\documentclass[aps,prd,onecolumn,showpacs,amsmath,amssymb]{revtex4-2}
\usepackage{graphicx}
\usepackage{subfig}
\usepackage{caption}
\usepackage{epstopdf}
\usepackage{color}
\usepackage{multirow}
\usepackage{setspace}
\usepackage{overpic}
\usepackage{amsmath}
\usepackage{amssymb}
\usepackage[bookmarksnumbered, pdfstartview=FitH,colorlinks,urlcolor=blue, citecolor=blue,linkcolor=blue] {hyperref}
\usepackage{lineno}
\usepackage{bm}
\usepackage{rotating}
\usepackage{verbatim}
\usepackage[utf8]{inputenc}
\graphicspath{{Fig/}}
\usepackage{bigstrut}
\usepackage{orcidlink}

\def\half{{1\over 2}}

\makeatletter
\renewcommand\tagform@[1]{\maketag@@@{(\ignorespaces#1\unskip\@@italiccorr)}}
\makeatother
\renewcommand{\eqref}[1]{(\ref{#1})}

\begin{document}


\title{Update analysis of $\psi(3686)\to p\bar{p}$}

\author{Zhi Gao$^1$\orcidlink{0009-0008-0493-0666}}
\email{zhi-gao@mail.nankai.edu.cn}
\author{Rong-Gang Ping$^{2}$\orcidlink{0000-0002-9577-4855}}\email{pingrg@ihep.ac.cn}
\author{Minggang Zhao$^1$\orcidlink{0000-0001-8785-6941}}\email{zhaomg@nankai.edu.cn}

\affiliation{$^1$ Nankai University, Tianjin 300071, China}
\affiliation{$^2$ Institute of High Energy Physics, Chinese Academy of Sciences, Beijing 100049, China}

\date{\today}

\begin{abstract}
  \rule{0ex}{3ex}
  We present an updated analysis of the angular distribution for $\psi(3686) \to p\bar{p}$ decay, taking into account transverse beam polarization, to investigate potential sources of forward-backward asymmetry and azimuthal modulation beyond the simple $1+\alpha\cos^2\theta$ form. We focus on the interference between the $\psi(3686)$ resonance and the two-photon exchange continuum process, as well as the background from initial-state-final-state radiation interference. A maximum-likelihood fit to the $\cos\theta$ distribution of $\psi(3686)\to p\bar{p}$ yields $\alpha = 1.00 \pm 0.03$, consistent with previous results. Our model predicts a significant $\cos(2\phi)$ modulation in the azimuthal angle, indicating the influence of transverse beam polarization. These findings motivate future two-dimensional angular analyses to fully disentangle the polarization and interference dynamics in charmonium decays to baryon pairs.

\end{abstract}

\maketitle

\section{Introduction}

Precise measurements of the electromagnetic form factors  of nucleons in the timelike region through processes such as $e^+e^- \to p\bar{p}$ and $e^+e^- \to n\bar{n}$ provide crucial insights into the non-perturbative structure of baryons. These form factors, encoding the spatial distributions of electric charge and magnetization, serve as fundamental probes of Quantum Chromodynamics (QCD) in the confinement regime. Data from the BESIII experiment \cite{BESIII:2019hdp,BESIII:2022rrg,Huang:2021xte}, with unprecedented precision, enable stringent tests of theoretical models and help resolve long-standing puzzles, such as the relative strength of photon couplings to protons and neutrons and the oscillatory behavior~\cite{BESIII:2015osc,BABAR:2013ppbar} observed in the effective form factors. Such studies bridge the spacelike and timelike domains, offering a deeper understanding of nucleon dynamics, the role of final-state interactions, and the transition between perturbative and non-perturbative QCD.

The differential cross section for the annihilation process $e^+e^- \to B\bar{B}$, where $B$ denotes a proton ($p$) or neutron ($n$), is expressed as a function of the center-of-mass energy squared $s$ and the baryon's polar angle $\theta$:
\begin{equation}
\frac{d\sigma_B(s)}{d\Omega} = \frac{\alpha_{\rm QED}^2 \beta C}{4s} \left[ |G_M(s)|^2 (1 + \cos^2\theta) + \frac{4m_B^2}{s} |G_E(s)|^2 \sin^2\theta \right],
\end{equation}
where $\alpha_{\rm QED}$ is the fine-structure constant, $\beta = \sqrt{1 - 4m_B^2/s}$ is the baryon velocity, $m_B$ is the baryon mass, and $C$ is the Coulomb enhancement factor for charged final states ($p\bar{p}$) or unity for neutral ones ($n\bar{n}$). The electric and magnetic Sachs form factors, $G_E(s)$ and $G_M(s)$, characterize the internal spatial distributions of charge and magnetization. Their ratio $|G_E/G_M|$ and individual moduli can be extracted by fitting the measured $\cos\theta$ distributions. In the timelike region ($s>0$), these form factors are complex, and their behavior near threshold provides critical information on final-state interactions and non-perturbative QCD dynamics

The angular distributions in $J/\psi \rightarrow p\bar{p}$ and $n\bar{n}$ decays, parameterized as $1+\alpha\cos^2\theta$, provide a sensitive probe of the underlying baryon structure and production dynamics \cite{Ping:2002uj,Kivel:2019wjh,Cieply:2004fe}. These decays proceed via three-gluon annihilation, offering a gluon-rich environment to test perturbative QCD and the role of gluon spin. The parameter $\alpha$ reflects the interplay between quark-mass effects, relativistic corrections, and baryon wave functions, distinguishing between asymptotic predictions and non-perturbative models. Measurements of $\alpha$ for both $p\bar{p}$ and $n\bar{n}$ final states can reveal isospin dependencies and SU(3) flavor-symmetry breaking, while comparisons with $\psi(3686)$ decays test heavy-quark spin symmetry and production mechanisms. Such studies are crucial for constraining quark models, understanding baryon formation, and exploring exotic hadronic configurations.

Existing data, primarily from BES and CLEO experiments, reveal that $\alpha_{p\bar{p}}$ for $\psi(3686)$ is consistently measured around $1.03\pm0.07$ \cite{BESIII:2018flj}, larger than the $J/\psi$ value of $0.676\pm0.055$ \cite{BES:2004vvp}, suggesting different dynamical origins. The first observation of $\psi(3686) \to n\bar{n}$ by BESIII yields $\alpha_{n\bar{n}}=0.68\pm0.16$ \cite{BESIII:2018flj}, which is consistent with $\alpha_{p\bar{p}}$ in $J/\psi$ decays within one standard deviation, but exhibits a discrepancy greater than $1\sigma$ compared to that from $\psi(3686)$ decays. The considerable uncertainties preclude definitive conclusions, thus motivate further precision studies to understand flavor-symmetry breaking and the role of quark-mass effects in charmonium decays into baryon pairs.

\begin{table}[h]
\centering
\caption{Measurements of branching fractions ($\mathcal{B}$) and angular distribution parameters ($\alpha$) for $\psi(3686) \to p\bar{p}$ and $\psi(3686) \to n\bar{n}$ decays. The first and second uncertainties are statistical and systematic, respectively.}
\label{tab:psi2s_branching}
\begin{tabular}{lllll}
\hline
\textbf{Source(Year)} & \textbf{Process} & \textbf{Branching Ratio $\mathcal{B}$} & \textbf{$\alpha$ parameter} & \textbf{Ref.} \\
\hline
BES (2004)            & $J/\psi \to p\bar{p}$          & $(2.26 \pm 0.01 \pm 0.14) \times 10^{-3}$ & $0.676 \pm 0.036 \pm 0.042$ & \cite{BES:2004vvp} \\
BES (2012)            & $J/\psi \to p\bar{p}$          & $(2.112 \pm 0.004 \pm 0.031) \times 10^{-3}$ & $0.595 \pm 0.012 \pm 0.015$ & \cite{BESIII:2012ion} \\
BES (2012)            & $J/\psi \to n\bar{n}$          & $(2.07 \pm 0.01 \pm 0.17) \times 10^{-3}$ & $0.50 \pm 0.04 \pm 0.21$ & \cite{BESIII:2012ion} \\
BES (2007)            & $\psi(3686) \to p\bar{p}$        & $(3.36 \pm 0.09 \pm 0.25) \times 10^{-4}$ & $0.85 \pm 0.24 \pm 0.04$     & \cite{BES:2006pax} \\
CLEO (2005)           & $\psi(3686) \to p\bar{p}$        & $(2.87 \pm 0.12 \pm 0.15) \times 10^{-4}$ & $\alpha$ not reported        & \cite{CLEO:2005ixz} \\
BESIII (2018)         & $\psi(3686) \to p\bar{p}$      & $(3.05 \pm 0.02 \pm 0.12) \times 10^{-4}$ & $1.03 \pm 0.06 \pm 0.03$     & \cite{BESIII:2018flj} \\
BESIII (2018)         & $\psi(3686) \to n\bar{n}$      & $(3.06 \pm 0.06 \pm 0.14) \times 10^{-4}$ & $0.68 \pm 0.12 \pm 0.11$     &\cite{BESIII:2018flj} \\
\hline
\end{tabular}
\end{table}

The measured branching ratios for $\psi(3686) \to p\bar{p}$ and $\psi(3686) \to n\bar{n}$ are consistent within one standard deviation \cite{BESIII:2018flj}, indicating negligible SU(3) flavor-breaking effects in the final state. This makes it particularly challenging to understand, within the quark model framework \cite{Ping:2002uj}, why the angular distribution parameter $\alpha$ for $\psi(3686) \to p\bar{p}$ differs so significantly from that for $\psi(3686) \to n\bar{n}$ and also deviates markedly from $\alpha$ in $J/\psi \to p\bar{p}$ decays \cite{BES:2004vvp}. Current experimental analyses are limited to one-dimensional (1D) angular distributions and do not account for potential asymmetry in the data \cite{BESIII:2018flj}. The accumulation of large $\psi(3686)$ data samples at BESIII call for an updated and more comprehensive measurement to clarify the underlying dynamics and to constrain theoretical models of baryon production in charmonium decays.

Extending the $1+\alpha\cos^2\theta$ parameterization of angular distribution in previous analyses, an improved measurement can be performed by analyzing the two-dimensional (2D) angular distribution in polar and azimuthal angles, incorporating the effects of transverse beam polarization in $e^+e^-$ collisions~\cite{ternov}. Several studies have investigated asymmetry phenomena in the electron-positron annihilation into protons process~\cite{Xia:2025rio,Czyz:2003ue}. To advance our understanding, a comprehensive approach incorporating potential physical asymmetries in the polar angular distribution is required. Accounting for asymmetries arising from interference contributions from two-photon processes and backgrounds from initial-final state radiation interference is essential for extracting precise angular coefficients and clarifying the underlying production dynamics.

\section{Analysis of angular distribution}\label{angHelAmp}
\begin{figure}[h!]
  \centering
  \subfloat[]{
  \begin{minipage}[t]{0.44\linewidth}
      \includegraphics[width=0.8\linewidth]{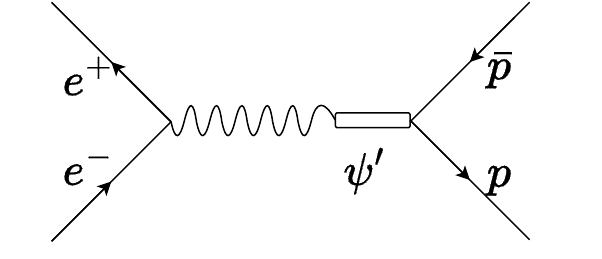}
  \end{minipage}
  }
  \subfloat[]{
  \begin{minipage}[t]{0.44\linewidth}
      \includegraphics[width=0.8\linewidth]{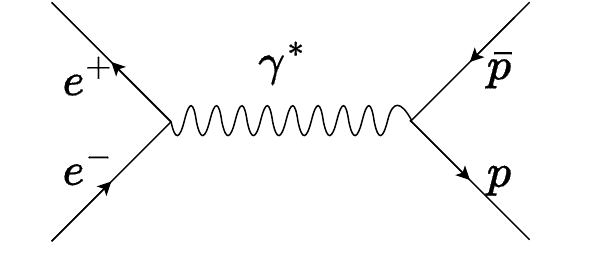}
  \end{minipage}
  }\\
  \subfloat[]{
  \begin{minipage}[t]{0.44\linewidth}
      \includegraphics[width=0.8\linewidth]{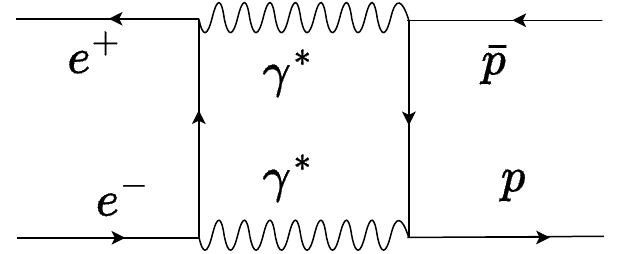}
  \end{minipage}
  }
  \subfloat[]{
  \begin{minipage}[t]{0.44\linewidth}
      \includegraphics[width=0.8\linewidth]{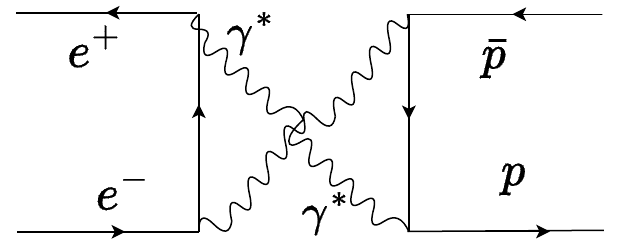}
  \end{minipage}
  }
  \caption{Feynman diagrams of one (a,b) and two (c,d) virtual photon processes $e^+e^-\to\gamma^*/\psi(3686)$ or $ \gamma^*\gamma^*\to p\bar{p}$}
  \label{fig::feynDia}
\end{figure}

At the $\psi(3686)$ energy point in electron-positron annihilation experiments such as those conducted at BEPCII, the final state \( p\bar{p} \) events can be produced through several processes, as illustrated in Fig. \ref{fig::feynDia}: resonant production via $\psi(3686)$, single-photon exchange, and two-photon exchange. Among these, the $\psi(3686)$ resonance dominates with a cross section of approximately 1700 nb, while the single-photon process provides a secondary contribution. The intermediate Z-boson exchange is neglected here due to its off-shell suppression, which leads to negligible production at this energy. Although the two-photon process appears at the order of \( \alpha_{\rm QED}^4 \) and is negligible, its interference with the $\psi(3686)$ amplitude could introduce non-trivial asymmetries in the polar angular distribution. The asymmetries can be tested experimentally using available data.

Experimentally, the \( p\bar{p} \) final state is characterized by two angular observables: the polar angle \( \theta \) and the azimuthal angle \( \phi \). The contributions from the aforementioned processes, including their coherent interference, can be described by a differential angular distribution that depends on both \( \theta \) and \( \phi \). This distribution provides a sensitive probe to disentangle the resonant, single-photon, and two-photon components, as well as to test potential interference effects that may arise from transverse beam polarization or higher-order QED contributions. Such analyses enhance the precision of cross-section measurements and enable deeper investigations into the dynamics of baryon pair production near charmonium resonances.

In this section, we formulate the angular distribution for the resonant and single-photon processes by incorporating the beam transverse polarization. Subsequently, we extend the analysis to include the interference with the two-photon exchange process. Finally, we formulate the interference terms from initial-state and final-state radiation.

\subsection{Beam transverse polarization}
At BEPCII, the transverse polarization of electron and positron beams arises naturally due to the Sokolov--Ternov effect \cite{ternov}---a self-polarization mechanism where spin-flip synchrotron radiation during circulation in the storage ring leads to a gradual buildup of polarization perpendicular to the beam direction. This process results in a characteristic polarization \(P_T\) that increases with storage time, reaching significant levels away from depolarization resonances. The spin density matrix of beams can be described by \cite{Cao:2024tvz}

\begin{align*}
\rho^- &= \frac{1}{2} \begin{pmatrix} 1 + P_z & P_T \\ P_T^* & 1 - P_z \end{pmatrix} \quad\text{for } e^-,\\
\rho^+ &= \frac{1}{2} \begin{pmatrix} 1 + \bar{P}_z & P_T \\ P_T^* & 1 - \bar{P}_z \end{pmatrix} \quad\text{for } e^+,
\end{align*}
where \(P_z\) and \(\tilde{P}_z\) denote the longitudinal polarizations of the electron and positron, respectively. In the symmetric BEPCII configuration with transverse polarization only, \(P_z = \tilde{P}_z = 0\) and \(P_T\) is non-zero, leading to off-diagonal elements that modify angular distributions and enable new observables in baryon decay analyses.

At the BEPCII experiment, the evolution of beam transverse polarization is intrinsically linked to the storage ring’s geometry and operational conditions. The polarization builds up naturally through the Sokolov–Ternov effect, with a characteristic time dependence described by \( P_t =\lvert P_T \rvert= P_0 (1 - e^{-t/\tau_0}) \), where \(\tau_0\) depends on the ring radius and beam energy. A simplified numerical estimate indicates that, after about one hour of beam injection at the \(\psi(3686)\) energy, the transverse polarization can reach approximately 28\% \cite{Cao:2024tvz}. This polarization can be experimentally measured offline using reference processes such as \(e^+e^-\to \mu^+\mu^-\) and \(\gamma\gamma\) production. The availability of transverse polarization introduces additional observables in the angular distribution of \(p\bar{p}\) final states, enabling a deeper investigation into the underlying production mechanisms—particularly the origin of azimuthal asymmetries that may arise from interference between resonant and continuum amplitudes.

\subsection{$e^+e^-\to\gamma^*/\psi(3686)\to p\bar{p}$}
For the process $e^+e^-\to p\bar p$ via the intermediate states $\gamma^*/\psi(3686)$, its angular distribution can be written as
\begin{equation}
 {d\sigma\over d\cos\theta} \propto \sum_{\substack{\lambda_i,\lambda'_i}} \rho^-_{\lambda_a,\lambda'_a}\rho^+_{\lambda_b,\lambda'_b} D^{1*}_{\lambda_a-\lambda_b,\lambda_c-\lambda_d}(\phi,\theta,0)D^{1}_{\lambda'_a-\lambda'_b,\lambda_c-\lambda_d}(\phi,\theta,0) T^{1}_{\lambda_a,\lambda_b,\lambda_c,\lambda_d}T^{1*}_{\lambda'_a,\lambda'_b,\lambda_c,\lambda_d},
\end{equation}
where $\lambda_a,~\lambda'_a=\pm 1/2$ and $\lambda_b,~\lambda'_b=\pm 1/2$ denote the helicity values of the electron and positron, respectively. $\lambda_c=\pm 1/2$ and $\lambda_d=\pm 1/2$ denote the helicities of the proton and antiproton. $D^J_{m_1,m_2}$ represents the Wigner-$D$ function in terms of the proton helicity angles $\theta$ and $\phi$,  and $T^{1}$  denotes the helicity amplitude.  The process conserves the parity, leading to the symmetry relation $T^{1}_{-\lambda_a,-\lambda_b,-\lambda_c,-\lambda_d}=T^{1}_{\lambda_a,\lambda_b,\lambda_c,\lambda_d}$.  Considering that the process \( e^+ e^- \to \gamma^* \) strictly obeys the helicity conservation rule, this leads to the helicity selection rule for \( e^+/e^- \), which can be expressed using the Kronecker delta \( \delta_{\lambda_a, -\lambda_b} \). After straightforward algebraic simplification, we obtain
\[
 {d\sigma\over d\cos\theta d\phi} \propto1 +\alpha \cos^2\theta + \alpha P_t^2 \sin^2\theta \cos 2\phi,
\]
where $\alpha$ is the angular distribution parameter for proton within the region $\alpha\in[-1,1]$.  Here the normalized relations,
$|T^{1}_{1/2,-1/2,1/2,-1/2}|^2$ $=(1+\alpha)/2$ and $|T^{1}_{1/2,-1/2,1/2,1/2}|^2=(1-\alpha)/4$, are used.

\subsection{Interference with two-photon process }
To account for the potentially asymmetric angular distribution of the proton, we consider the interference between processes involving \(\psi(3686)\) and two-photon contributions. The proton distribution from \(\psi(3686)\) decay yields a symmetric profile, corresponding to the \(S\)- and \(D\)-wave components of the proton, whereas the two-photon process also comprises contributions from partial waves with $L=1,3$. Interference between these terms therefore leads to an asymmetric distribution. In this analysis only \(P\)-wave contribution is considered. The amplitudes for processes (a) and (c) in Fig. \ref{fig::feynDia} can be uniformly expressed as
\begin{equation}
\mathcal{A}_{a/c}(\lambda_a,\lambda_b,\lambda_c,\lambda_d)=  {1\over 4\pi}\sum_{\substack{J_{a/c}}} (2J_{a/c}+1) D^{J_{a/c}*}_{\lambda_{a}-\lambda_{b},\lambda_{c}-\lambda_{d}}(\phi,\theta,0)T^{J_{a/c}}_{\lambda_{a},\lambda_{b},\lambda_{c},\lambda_{d}},
\end{equation}
where $J_{a/c}$ represents the spin of the intermediate state. For the $\psi(3686)$ process, we take $J_a = 1$. In the case of the two-photon process, there are three possible spin configurations: $J_c = 0, 1, 2$. The helicity selection rule forbids simultaneous coupling of one and two photons of $J_c=0$ system to the $e^+e^-$ beams, which cancels the interference contribution. For the $J_c = 1$ configuration, $C$-parity conservation results in only one non-zero helicity amplitude, $T_{1/2,-1/2,1/2,-1/2}$, yielding a symmetric distribution of order \( \alpha_{\rm QED}^4 \), which can be neglected. Thus,  we only consider the interference term of the case $J_c = 2$ with resonance, at level of $\alpha_{QED}^2$ order.  Then we have the interference contribution $\mathcal{I}_{ac} \propto  \sum_{\substack{\lambda_i}}  \rho^-_{\lambda_a,\lambda'_a}\rho^+_{\lambda_b,\lambda'_b} Re(\mathcal{A}_a\mathcal{A}^*_c)$, and
\begin{eqnarray}
    \mathcal{A}_a\mathcal{A}^*_c &\propto& D^{1*}_{\lambda_a-\lambda_b,\lambda_c-\lambda_d}(\phi,\theta,0)D^{2}_{\lambda'_a-\lambda'_b,\lambda_c-\lambda_d}(\phi,\theta,0) T^{1}_{\lambda_a,\lambda_b,\lambda_c,\lambda_d}T^{2*}_{\lambda'_a,\lambda'_b,\lambda_c,\lambda_d}.
\end{eqnarray}
Using the parity-conserved relations, $T^{2}_{-\lambda_a,-\lambda_b,-\lambda_c,-\lambda_d}=T^{2}_{\lambda_a,\lambda_b,\lambda_c,\lambda_d}$,  one has the reduced formula:
\begin{eqnarray}
        {d\mathcal{I}_{ac}\over d\cos\theta d\phi} &\propto& \cos\theta \biggl[ A \cdot [ \cos^2\theta + P_t^2 \cos(2\phi) \sin^2\theta ] - B \cdot \sin^2\theta \biggr],\nonumber\textrm{~with}\\
        A&=&8\sqrt{2}\sqrt{1+\alpha} \cos\left(\frac{1+\alpha}{2} + \beta\right) H^2_{\half,-\half,\half,-\half},\\
        B&=&8\sqrt{3}\sqrt{| \alpha - 1|} \left[ -1 + P_t^2 \cos(2\phi) \right] \cos\left(\frac{3}{2}(\alpha - 1)\right) H^2_{\half,-\half,\half,\half},
\end{eqnarray}
where $\beta$ is the phase angle difference between $T^{1}_{1/2,-1/2,1/2,1/2}$ and $T^{1}_{1/2,-1/2,1/2,-1/2}$, and $\alpha$ is the angular distribution parameter of proton.  Here, we treat \( H^2 \) as a real number, based on the consideration that the photon coupling with charged particles via \( (ie)^2 \) , and the fact that both the photon and proton are zero-width particles leading to a real amplitude. A numerical investigation in Ref. \cite{Chen:2008hka} showed that the form factor for photon–proton coupling has a negligible imaginary part, which justifies the use of a real-amplitude approximation.

\begin{figure}[h!]
\centering
\subfloat[]{
\begin{minipage}[b]{0.44\linewidth}
  \centering
   \includegraphics[width=0.9\linewidth]{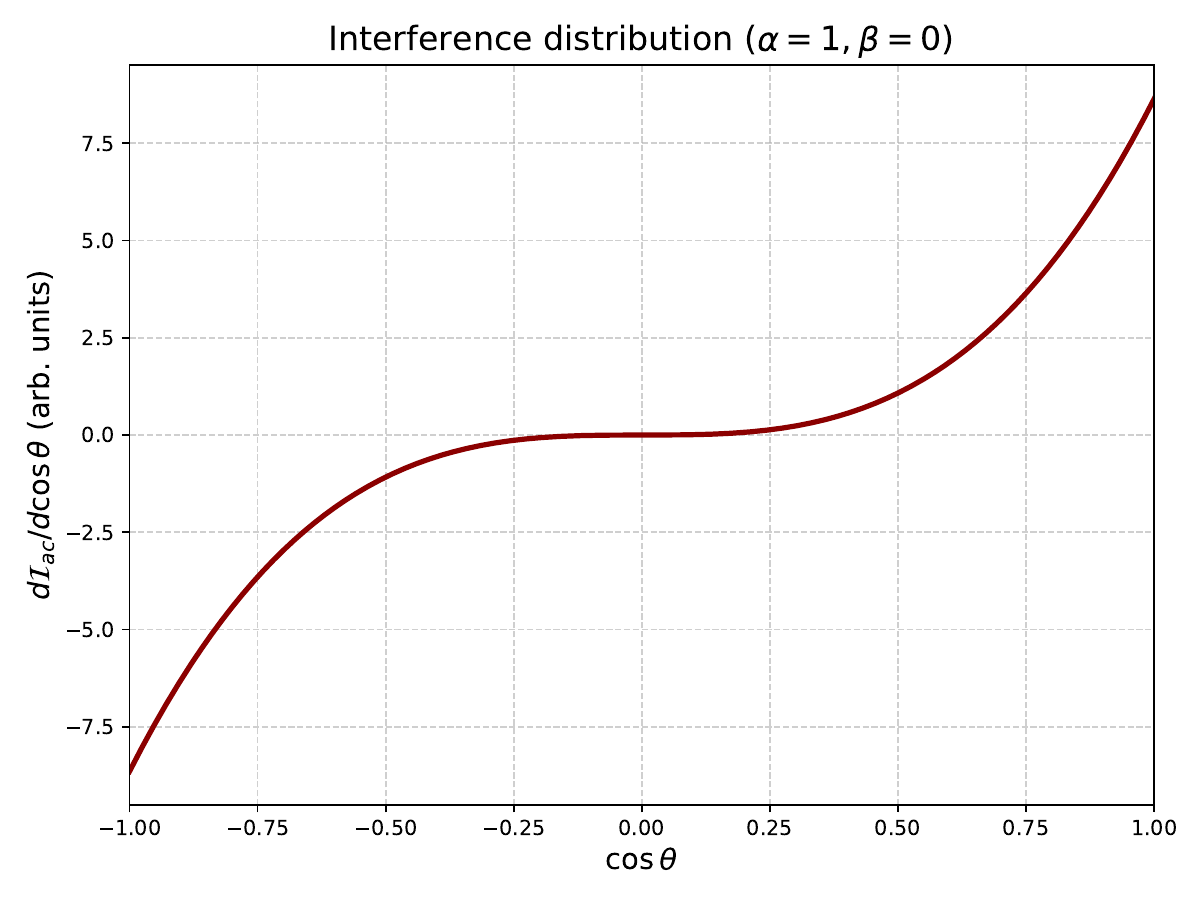}
\end{minipage} }
  \caption{Asymmetric proton angular distribution resulting from the interference between resonance and two-photon processes.}
  \label{fig::g2ggint}
\end{figure}

If the $\phi$ is integrated out, one has a reduced distribution \cite{Chen:2008hka,Xia:2025rio}
\begin{eqnarray}
\frac{d\mathcal{I}_{ac}}{d\cos\theta} &\propto& B \cos\theta + (A - B) \cos^3\theta \nonumber, \text{~with}\\
A &=& 8\sqrt{2}\sqrt{1+\alpha} \cos\!\left(\frac{1+\alpha}{2} + \beta\right) H^2_{\frac12,-\frac12,\frac12,-\frac12},\\
B &=& 8\sqrt{3}\sqrt{| \alpha - 1|} \cos\!\left(\frac{3}{2}(\alpha - 1)\right) H^2_{\frac12,-\frac12,\frac12,\frac12}.\nonumber
\end{eqnarray}
The asymmetry distribution is plotted in Fig. \ref{fig::g2ggint} for the parameter choice $\alpha = 1, \beta = 0$, and is displayed in arbitrary units.

\subsection{ISR-FSR interference}\label{ISRFSRint}

The corrections from initial-state radiation (ISR) and final-state radiation (FSR) have been extensively studied in quantum electrodynamics. Calculations show that the photon radiation from charged particles is concentrated near the direction of the particle's motion, decreases rapidly as the angle increases, and the radiation probability is inversely proportional to the square of the charged particle's mass. The higher the energy of the radiated photon, the lower the probability of its emission.

\begin{figure}[h!]
  \centering
	\subfloat[]{
		\begin{minipage}[b]{0.44\linewidth}
  \centering
      \includegraphics[width=0.9\linewidth]{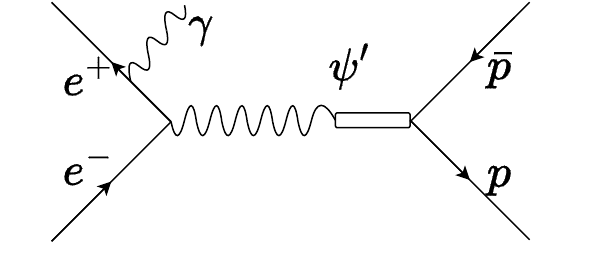}
	\end{minipage} }
	\subfloat[]{
		\begin{minipage}[b]{0.44\linewidth}
  \centering
      \includegraphics[width=0.9\linewidth]{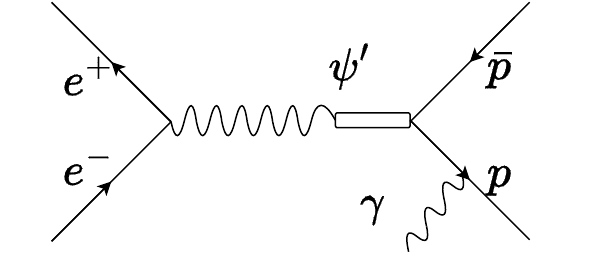}
	\end{minipage} }
  \caption{Tree level of Feynman diagrams for ISR (a), FSR (b) processes $e^+e^-\to\psi(3686)\to p\bar{p}$.}
  \label{fig::feynDiaIFSRint}
  \end{figure}

In ISR, the photons produced are mostly soft photons along the beam direction. In FSR, the photons radiated by protons and antiprotons are mainly emitted along their respective directions of motion. Their energies are mostly below the sensitivity threshold of photon detectors, so they may not be detected. Such undetected final-state radiation processes have the same final observed state as the signal channel, thereby forming an important type of background. The tree level of ISR-FSR diagram is shown in Fig. \ref{fig::feynDiaIFSRint}.

Although the probability of FSR from protons (antiprotons) is much smaller than that of ISR, the interference between them can lead to an asymmetry in the angular distribution. This is because, in ISR, the \( p\bar{p} \) system has a parity of \( 1^- \) and is in a superposition of \( S \)-wave and \( D \)-wave states, while in FSR, the \( p\bar{p} \) system is in a \( P \)-wave state. The interference between states of different parity can produce an asymmetric angular distribution. This asymmetric profile has been found for the process \cite{Czyz:2003ue}.

To investigate the contribution of the ISR–FSR interference to the asymmetry, we compute the interference term between diagrams (a) and (b) in Fig.~\ref{fig::feynDiaIFSRint}, which can be written as \(I_{12} \propto 2\,\mathcal{R}e\bigl(\rho^+\rho^- \mathcal{M}_1 \mathcal{M}_2^*\bigr)\). Here \(\rho^+\) and \(\rho^-\) denote the spin density matrices of the positron and electron, respectively, while \(\mathcal{M}_1\) and \(\mathcal{M}_2\) represent the amplitudes corresponding to diagrams (a) and (b) in Fig.~\ref{fig::feynDiaIFSRint}.

For diagram (a) the amplitude reads
\begin{equation}
\begin{split}
\mathcal{M}_1 \sim
      &D_{\lambda_+' - \lambda_-,\; \lambda_p - \lambda_{\bar{p}}}^1(\phi, \theta, 0)\,
       D_{\lambda_+',\; \lambda_+ - \lambda_\gamma}^{\frac12}(\phi_1, \theta_1, 0) \\
      & \times H_{\lambda'_+,\lambda_-,\lambda_p,\lambda_{\bar{p}}}\,
        G_{\lambda_+, \lambda_\gamma},
\end{split}
\end{equation}
where \(\lambda_+\), \(\lambda_-\), \(\lambda_p\), and \(\lambda_{\bar{p}}\) are the helicities of the positron, electron, proton, and antiproton, respectively. The factor \(G_{\lambda_+,\lambda_\gamma}\) is the helicity amplitude for the subprocess \(e^+\to \gamma e^+\), described by the helicity angles \((\phi_1,\theta_1)\). The quantity \(H_{\lambda'_+,\lambda_-,\lambda_p,\lambda_{\bar{p}}}\) denotes the helicity amplitude for the coupling of the \(e^+e^-\) pair to the \(p\bar{p}\) system.

For diagram (b) the amplitude is
\begin{equation}
\begin{split}
\mathcal{M}_2 \sim
      &D_{\lambda_+ - \lambda_-,\; \lambda_p' - \lambda_{\bar{p}}}^1(\phi, \theta, 0)\,
       D_{\lambda_p',\; \lambda_p - \lambda_\gamma}^{\frac12}(\phi_2, \theta_2, 0) \\
      & \times H_{\lambda_+,\lambda_-,\lambda'_p,\lambda_{\bar{p}}}\,
        F_{\lambda_p, \lambda_\gamma},
\end{split}
\end{equation}
where \(F_{\lambda_p,\lambda_\gamma}\) is the helicity amplitude for the radiation \(p\to\gamma p\) with helicity angles \((\phi_2,\theta_2)\), and \(H_{\lambda_+,\lambda_-,\lambda'_p,\lambda_{\bar{p}}}\) again describes the \(e^+e^- \to p\bar{p}\) vertex.

Taking into account both ISR and the two FSR possibilities for the cases the photon emitted from either the proton or the antiproton, the interference contribution to the cross section can be expressed as
\begin{equation}
\sigma_{\rm ISR-FSR} \propto
2\,\operatorname{Re}\!\Bigl[
\rho^+\rho^- \mathcal{M}_1
\bigl(\mathcal{M}_2^* + \mathcal{M}_2^*([\theta_2,\phi_2]\to[\theta_3,\phi_3])\bigr)
\Bigr],
\end{equation}
where the replacement \([\theta_2,\phi_2]\to[\theta_3,\phi_3]\) indicates that the FSR photon is emitted from the antiproton, with corresponding helicity angles \(\theta_3,\phi_3\). Events in which the ISR or FSR photon escapes detection (because it is soft or collinear) lead to the same visible final state as the signal channel \(e^+e^-\to p\bar{p}\). Such configurations therefore constitute an irreducible background, which we treat by integrating over the unobserved photon angles.

If the angles for the ISR(FSR) photons are integrated out,  then one has the angular distribution,
\begin{equation}
\begin{split}
    &\frac{d {\sigma}_{\rm ISR-FSR}}{d \cos \theta \, d \phi} \propto\\
    &  4 \left( 6 + 6 \cos (2 \theta) - \sqrt{2} \sin \theta + 6 P_t^2 \cos (2 \phi) \sin \theta \left( \sqrt{2} \sin^2 \left( \frac{\theta}{2} \right) + 2 \sin \theta \right) \right. \\
    & \left. + \cos \theta \left( 4 + 3 \sqrt{2} \sin \theta \right) \right) \alpha - 2 \left( 4 \sqrt{2} \cos \theta + \sqrt{2} \cos (2 \theta) \right.\\
    &\left. + 2 \sin \theta \left( 12 + \sqrt{2} P_t^2 \cos (2 \phi) \sin \theta \right) \right) \sqrt{1 - \alpha^2} \\
    & + 2 \left( 24 + 8 \cos \theta + 8 \sqrt{2} P_t^2 \cos \left( \frac{\theta}{2} \right) \cos (2 \phi) \sin^3 \left( \frac{\theta}{2} \right) \right.\\
    &- 6 \sqrt{2} \sin \theta  \left. + \sqrt{2} \sin (2 \theta) + 5 \sqrt{2 - 2 \alpha^2} \right) ,
\end{split}
\end{equation}
where $P_t$ and $\alpha$ are the beam transverse polarization and the angular distribution parameter of proton, respectively.

If the $\phi$ is integrated out, one has
\begin{equation}
\begin{split}
\frac{d\sigma_{\rm ISR-FSR}}{d\cos\theta} &\propto 4 \Bigl( 6 + 6\cos(2\theta) - \sqrt{2}\sin\theta + 4\cos\theta + 3\sqrt{2}\sin\theta\cos\theta \Bigr) \alpha \\
&- 2 \Bigl( 4\sqrt{2}\cos\theta + \sqrt{2}\cos(2\theta) + 24\sin\theta \Bigr) \sqrt{1 - \alpha^2} \\
&+ 2 \Bigl( 24 + 8\cos\theta - 6\sqrt{2}\sin\theta + \sqrt{2}\sin(2\theta) + 5\sqrt{2}\sqrt{1 - \alpha^2} \Bigr).
\end{split}
\end{equation}

\begin{figure}[h!]
\centering
\subfloat[]{
\begin{minipage}[b]{0.44\linewidth}
  \centering
   \includegraphics[width=0.9\linewidth]{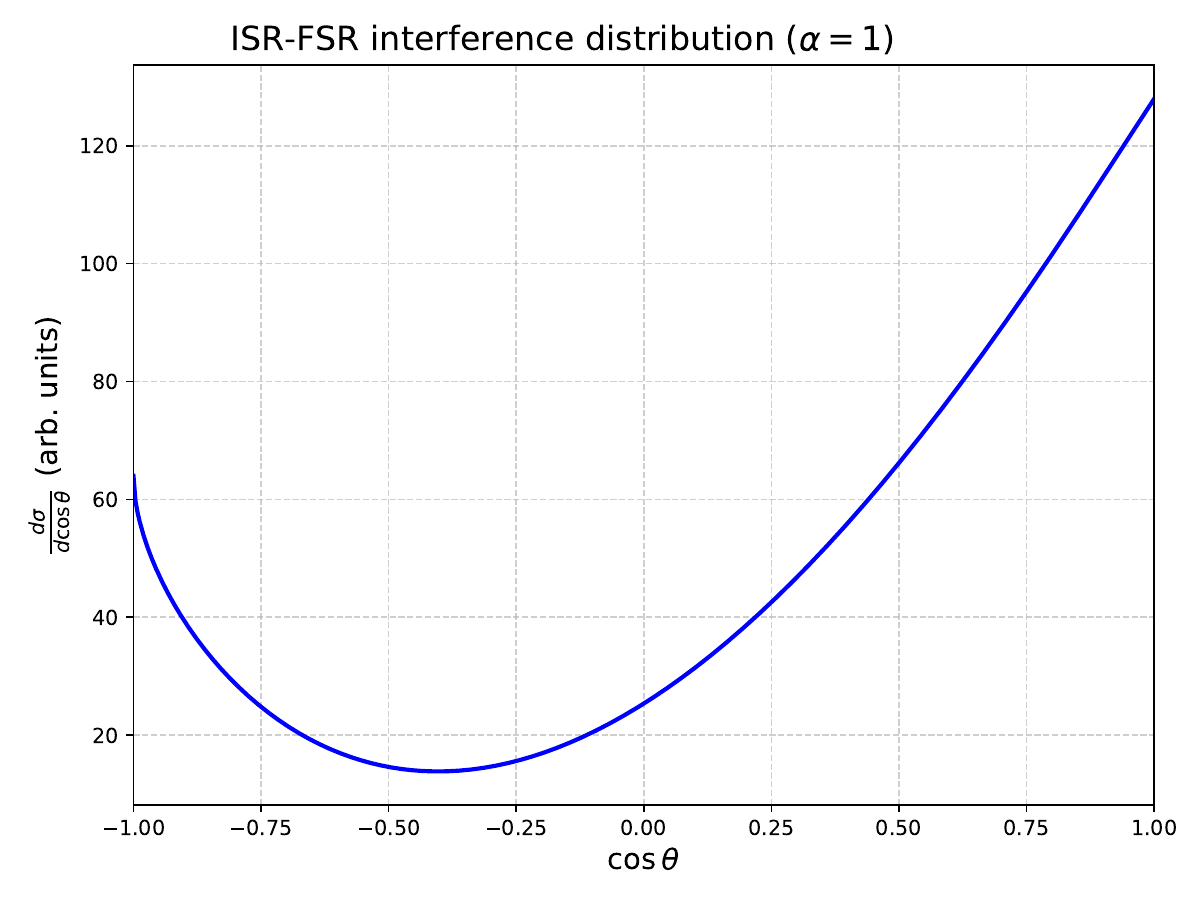}
\end{minipage} }
  \caption{ Asymmetric proton angular distribution resulting from the interference between ISR and FSR processes.}
  \label{fig::ISRFSR}
\end{figure}

\section{Update analysis of $e^+e^-\to p\bar p$}
\subsection{Updating the fit to the measured proton $\cos\theta$ distribution}

The angular distributions for $\psi(3686) \to p\bar{p}$ and $n\bar{n}$ were previously measured by the BESIII collaboration~\cite{BESIII:2018flj}, using $1.07 \times 10^8$ $\psi(3686)$ events. In that analysis, the angular distribution was described by $1+\alpha\cos^2\theta$. They explored the possibility of a forward-backward asymmetry by introducing an odd $\cos\theta$ contribution into the angular distribution, parametrized as $1+\beta\cos\theta+\alpha\cos^2\theta$. In their analysis, while the anomalous term was found to be consistent with zero, $\beta_{n\bar{n}}=0.04\pm 0.05$ and $\beta_{p\bar{p}}=0.01\pm0.02$, its physical origin remained unspecified and the treatment should be understood beyond the empirical asymmetry. Moreover, they did not account for the background from ISR-FSR interference. In order to properly account for the two-photon interference asymmetry and incorporate the ISR–FSR interference background, we performed a renewed fit to the angular distribution incorporating the asymmetry sources.

The fitting is carried out using the maximum likelihood (ML) method. The likelihood function $\mathcal{L}$ is defined as:
\begin{equation}
    \mathcal{L}(\vec{p}) = \prod_{i=1}^{n} P\left(y_i \mid \frac{d\sigma^{\rm tot}}{d\cos\theta}(\theta_i;\vec{p}), \sigma_{i}^{-}, \sigma_{i}^{+}\right),
\end{equation}
where $y_i$ is the observed number of events at $\theta_i$, $\frac{d\sigma^{\rm tot}}{d\cos\theta}(\theta_i;\vec{p})$ is the theoretical prediction evaluated with parameters $\vec{p}$, and $\sigma_{i}^{-}$ and $\sigma_{i}^{+}$ are the lower and upper uncertainties of the measurement, respectively. The probability density function $P$ is constructed by a Gaussian function.

The fitting procedure obtains the optimal parameter estimates by minimizing the negative log-likelihood:
\begin{equation}
    -\ln\mathcal{L}(\vec{p}) = -\sum_{i=1}^{n} \ln P\left(y_i \mid \frac{d\sigma^{\rm tot}}{d\cos\theta}(\theta_i;\vec{p}), \sigma_{i}^{-}, \sigma_{i}^{+}\right).
\end{equation}

Since only a 1D distribution in $\cos\theta$ is available in the published analysis~\cite{BESIII:2018flj} , the theoretical angular distribution function $\frac{d\sigma^{\rm tot}}{d\cos\theta}$ is defined by integrating out $\varphi$:
\begin{equation}
    \frac{d\sigma_p^{\rm tot}}{d\cos\theta} = N\epsilon(\theta)\left(\frac{d\sigma}{d\cos\theta} + \frac{d\mathcal{I}_{ac}}{d\cos\theta} + c_1 \frac{d\sigma_{\rm ISR-FSR}}{d\cos\theta}\right),
\end{equation}
here $N$ is a normalization factor, $\epsilon(\theta)$ represents the efficiency correction curve adopted from the experimental analysis~\cite{BESIII:2018flj}, and $c_1$ is a fit parameter scaling the contribution from initial--final state radiation (ISR--FSR).

\begin{figure}[!h]
	\centering
\subfloat[]{
\begin{minipage}[b]{0.44\linewidth}
  \centering
   \includegraphics[width=0.9\linewidth]{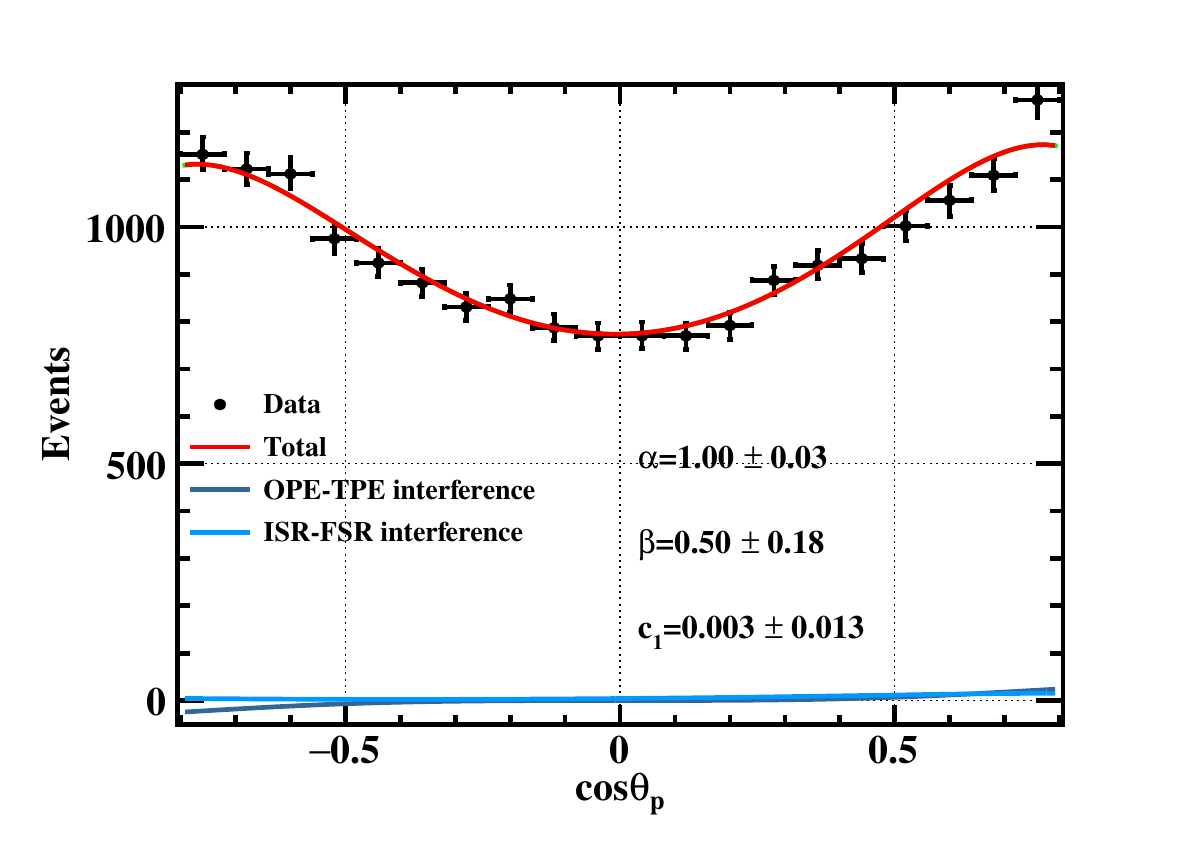}
\end{minipage} }
\subfloat[]{
\begin{minipage}[b]{0.44\linewidth}
  \centering
   \includegraphics[width=0.9\linewidth]{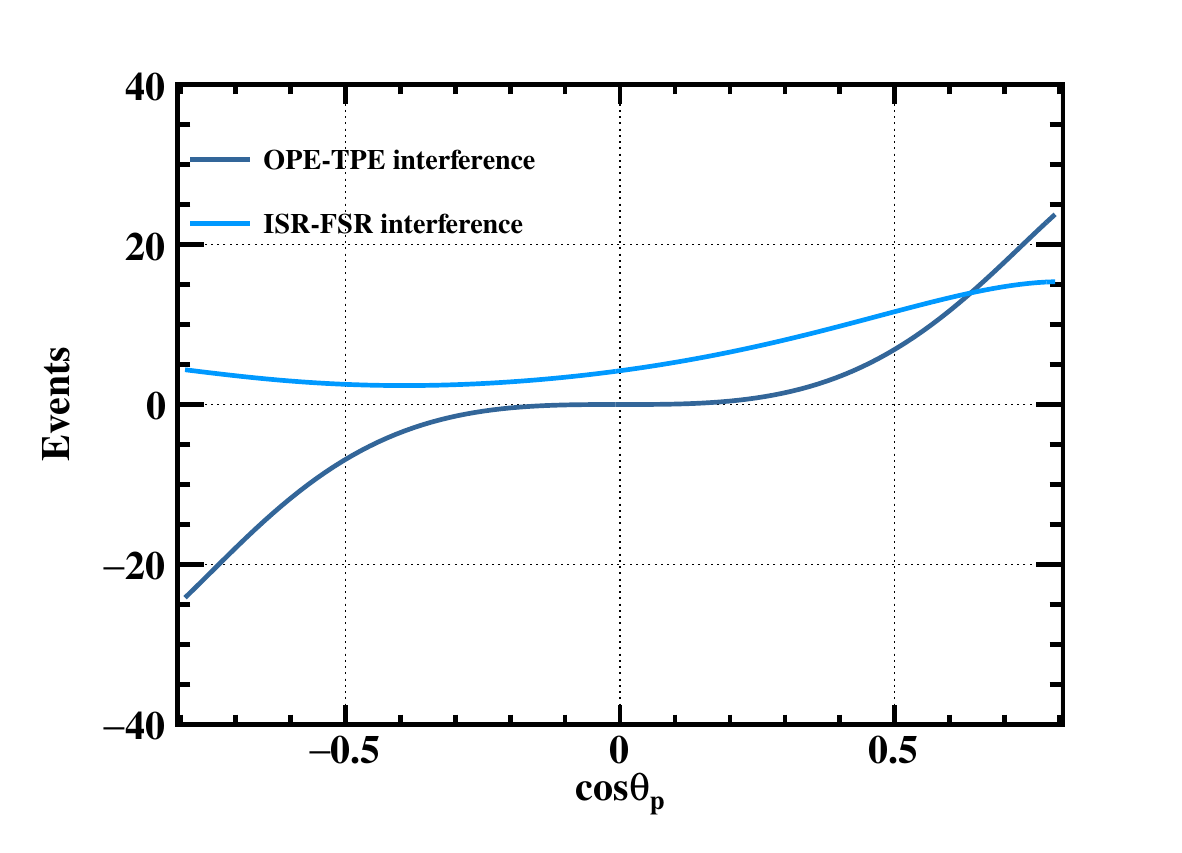}
\end{minipage} }
	\caption{The plot of $\cos\theta_{p}$ distribution. The contribution of the interference with two-photon exchange process is shown by a azure line, the contribution of initial state radiation and final state radiation interference is shown by a light blue line. The total PDF is shown by a red line. (b) shows the detail of distributions of the interference terms.}
	\label{fig::AngularfitPubforppbar}
\end{figure}

The angular parameter $\alpha$ was measured to be $1.03 \pm 0.06 \pm 0.03$~\cite{BESIII:2018flj} with the $1+\alpha\cos^2\theta$ parameterization. We now perform a new ML fit to this reported angular distribution to test the validity of our proposed formalism. During the ML fit, we treat $H^2_{\frac12,-\frac12,\frac12,-\frac12}$ and $H^2_{\frac12,-\frac12,\frac12,\frac12}$ as real parameters and float $H^2_{\frac12,-\frac12,\frac12,-\frac12}$, $H^2_{\frac12,-\frac12,\frac12,\frac12}$, $N$, $c_1$, $
\alpha$, and $\beta$. The results are shown in Fig. \ref{fig::AngularfitPubforppbar}, and the fitted values of main parameters are summarized in Table. \ref{tab::fitresultv}.

\begin{table}[!h]
  \centering
  \caption{Result of parameters of the proton angular distribution fit.}
  \begin{tabular}{cc}
    \hline\hline
Parameter  &  Value \bigstrut \\ \hline
$\alpha$ & $ 1.00 \pm 0.03 $     \\ \hline
$\beta$  &  $ 0.50 \pm 0.18 $   \\ \hline
 $ H_{1/2,-1/2,1/2,-1/2}^2$   &   $0.058\pm 0.072$                  \\ \hline
  $ H_{1/2,-1/2,1/2,1/2}^2$   &   $0.00  \pm 0.01 $      \bigstrut\\
    \hline\hline
  \end{tabular}
  \label{tab::fitresultv}
\end{table}

\begin{figure}[htb]
	\centering	
\subfloat[]{
\begin{minipage}[b]{0.44\linewidth}
  \centering
   \includegraphics[width=0.9\linewidth]{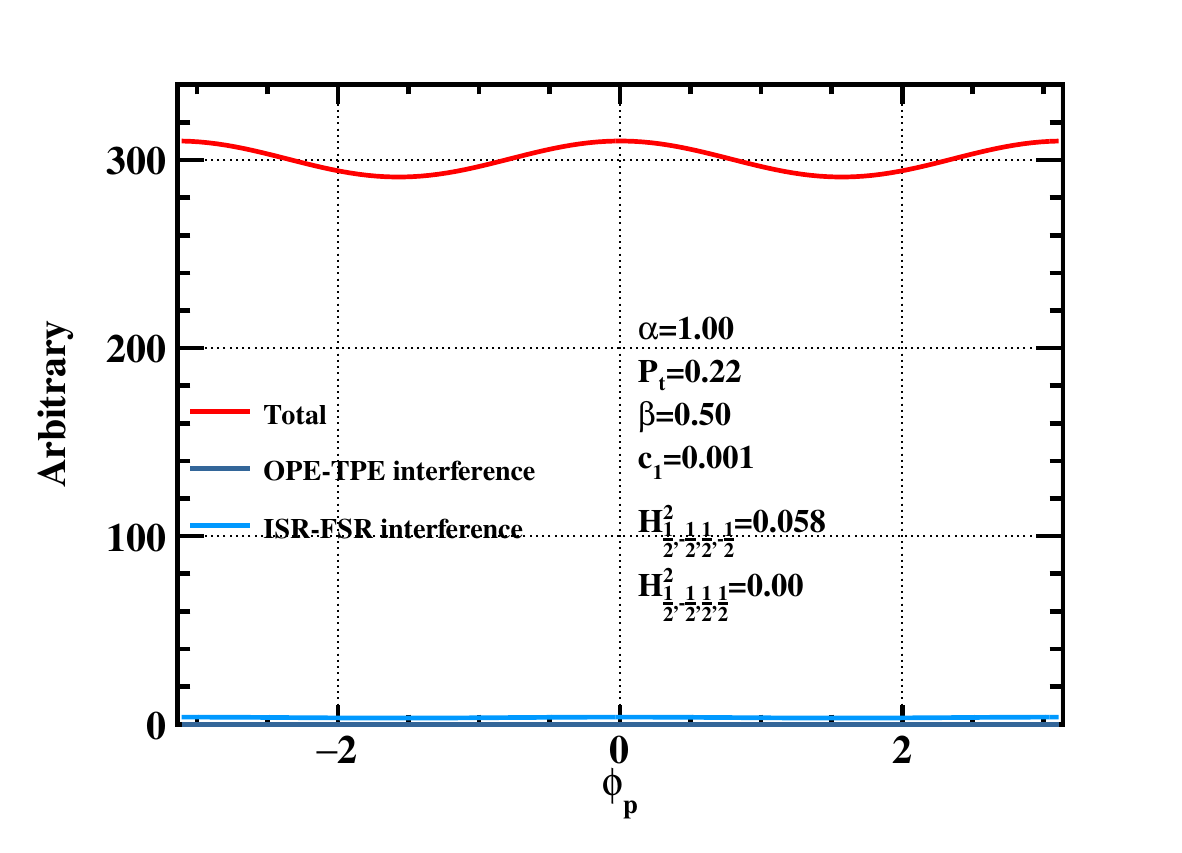}
\end{minipage} }
\subfloat[]{
\begin{minipage}[b]{0.44\linewidth}
  \centering
   \includegraphics[width=0.9\linewidth]{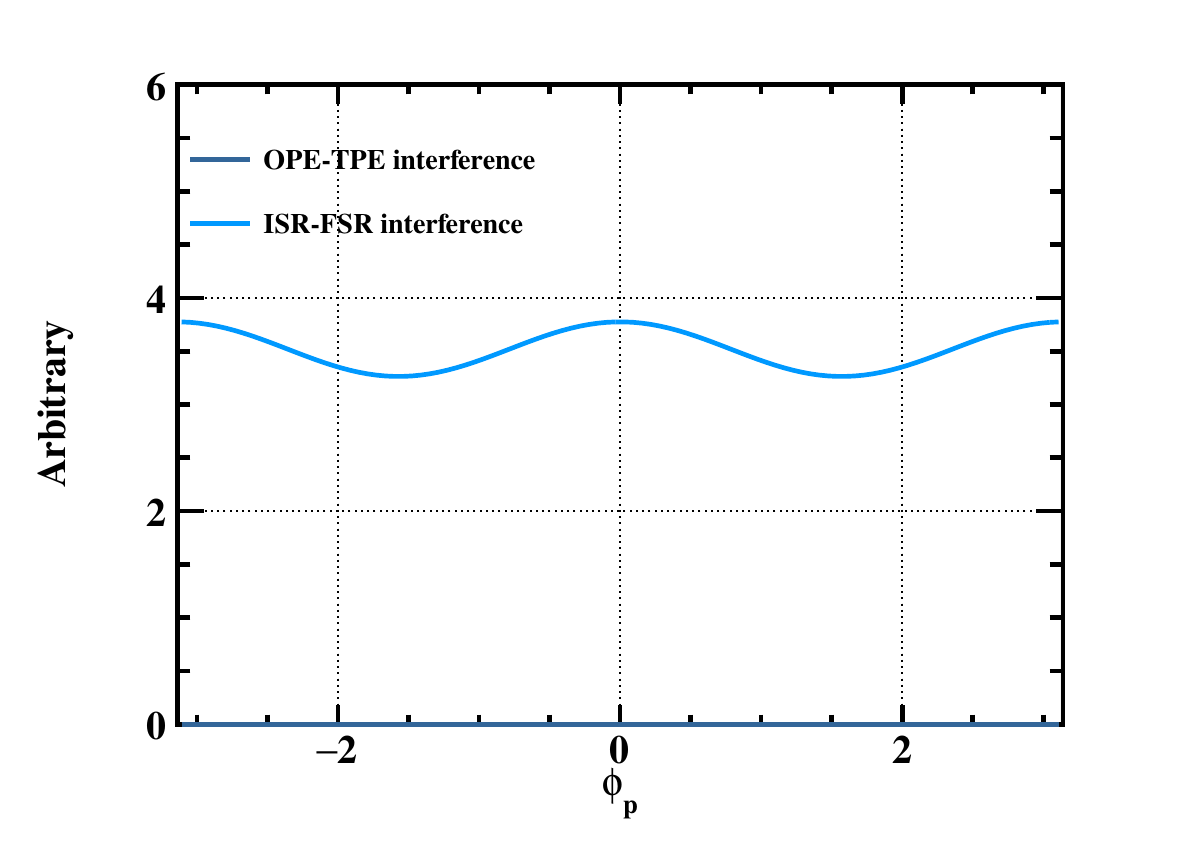}
\end{minipage} }
	\caption{The plot of expected $\phi_{p}$ distribution at $\alpha=1.00$, $P_t=0.22$, $\beta=0.50$, $c_1=0.003$, $H_{1/2,-1/2,1/2,-1/2}^2=0.058$, and $H_{1/2,-1/2,1/2,1/2}^2=0.0$. (b) shows the detail of distributions of the interference terms.}
	\label{fig::Expphidisforppbar}
\end{figure}

Our fitted angular parameter for the proton, $\alpha = 1.00 \pm 0.03$, is consistent with the previously published value of $\alpha = 1.03 \pm 0.06 \pm 0.03$ within one standard deviation. With the fitted parameters, we plot the $\phi$ distribution as shown in Fig. \ref{fig::Expphidisforppbar}.
A distinct feature is that the $\psi(3686)$ production distribution deviates from the phase space distribution, exhibiting a $\sin(2\phi)$ modulation, while the phase space is flat. The interference contributions from the resonance and the two-photon process are relatively small but not trivial. In contrast, the ISR–FSR interference background is negligible.  In order to uncover transverse polarization and the interference behavior between the resonance and two-photon processes, a 2D analysis of the nucleon's polar and azimuthal angles should be performed. We hope this approach can be realized in future data analysis.

\subsection{Sensitivity of $P_t$ measurement}

To evaluate the sensitivity of $P_t$ measurement, we generate 0.17 million toy Monte Carlo (MC) events using the amplitude model for the decay $\psi(3686)\to p\bar{p}$, at each of the $P_t$ values: $P_t=0.1,~0.2,~0.4,~0.6,~0.8$. Other parameters are fixed to $\alpha=1.00$, $\beta=0.50$, $c_1=0.003$, $H_{1/2,-1/2,1/2,-1/2}^2=0.058$, and $H_{1/2,-1/2,1/2,1/2}^2=0.0$. For estimating the sensitivity versus the number of events, multiple samples are prepared by  sampling randomly from the the 0.17 million toy MC events. Additionally, about 0.38 million phase space MC events are generated for calculating the normalization factor in the likelihood function.

The 2D differential cross section is defined as:
\begin{equation}
    \frac{d\sigma^{\rm tot}}{d\cos\theta d\varphi} = \frac{d\sigma}{d\cos\theta d\varphi} + \frac{d\mathcal{I}_{ac}}{d\cos\theta d\varphi} + c_1 \frac{d\sigma_{\rm ISR-FSR}}{d\cos\theta d\varphi},
\end{equation}

The likelihood function for $i$-th event is defined as:
\begin{equation}
  \mathcal{L}^i(\cos\theta_i, \phi_i| \alpha, P_t, \beta, c_1, H^2_{\frac12,-\frac12,\frac12,-\frac12}, H^2_{\frac12,-\frac12,\frac12,\frac12}) = \frac{d\sigma^{\rm tot}/d\cos\theta_i d\varphi_i}{\sigma^\text{tot}},
\end{equation}
here the normalization factor $\sigma^\text{tot}$ is calculated with the MC method, namely $\sigma^\text{tot}=\frac{1}{N_{\rm MC}}\sum_{i}^{N_{\rm MC}}{d\sigma^{\rm tot}\over d\cos\theta_i d\varphi_i}$, here $N_{\rm MC}$ is the number of phase space MC events. Instead, we minimize the object function for the $N_\text{dt}$ events
\begin{equation}\label{eq::minimizef}
  -lnL=-\sum_{i}^{N_{\rm dt}}ln\mathcal{L}^i,
\end{equation}
using the TMinuit package~\cite{Minuit}. The statistical significance to measure the $P_t$ signal is estimated by comparing the differences in the log-likelihood values ($\Delta(-lnL)$) with and without $P_t$ parameter.  The resulting significance for $P_t$ parameter is estimated by $\sqrt{|\Delta(-lnL)|}$, and it is then plotted as a function of the number of events for varying $P_t$ values.
\begin{figure}[h!]
  \centering
      \includegraphics[width=0.44\linewidth]{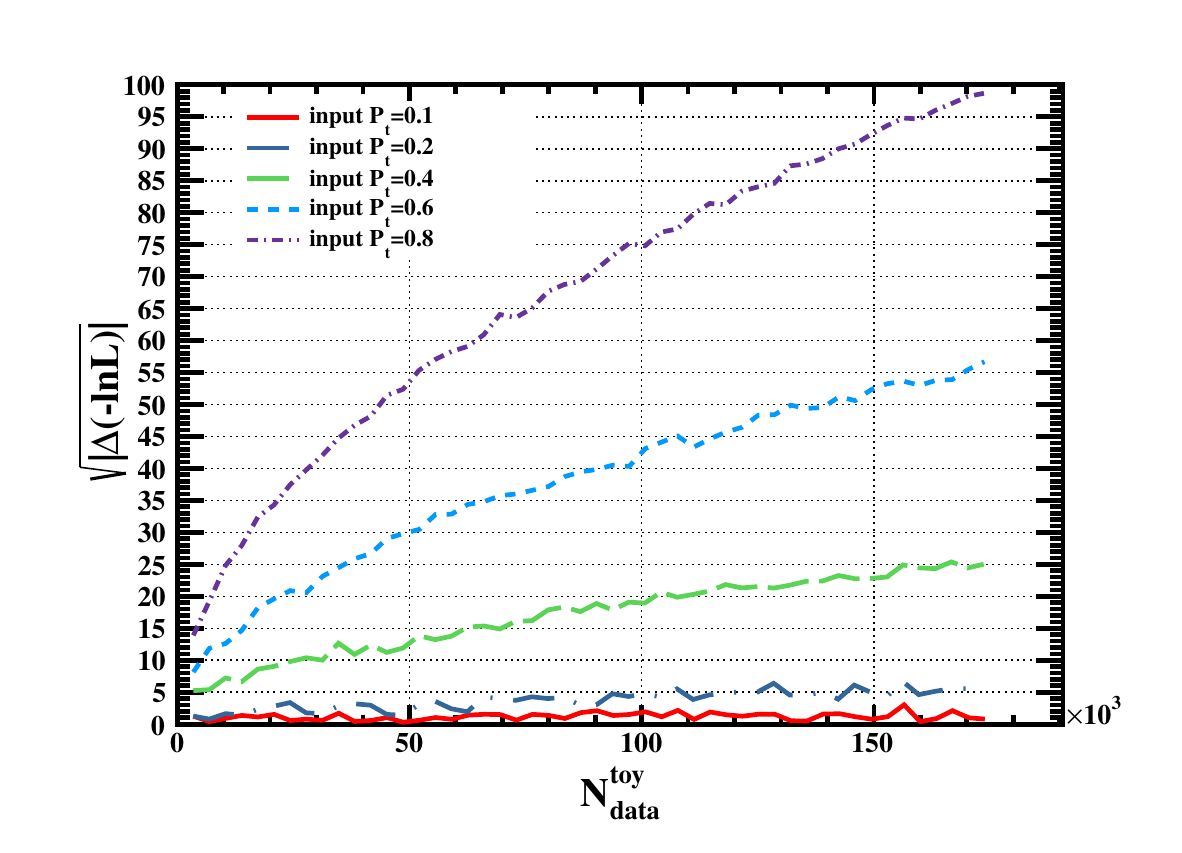}
  \caption{The plot of $P_t$ significance as a function of the number of toy sample events and the input $P_t$ value.}
  \label{fig::Ptsensitivity}
\end{figure}

As shown in Fig. \ref{fig::Ptsensitivity}, a significance of $5\sigma$ can be reached with only about 0.1 million events when $P_t \geq 0.2$. Given that BESIII has accumulated 2712.4 million $\psi(3686)$ events, approximately 0.5 million $\psi(3686)\to p\bar{p}$ events can be selected. This sample allows us to precisely measure the transverse beam polarization at BEPCII, which is theoretically expected to reach about 0.28~\cite{Cao:2024tvz} after one hour of beam energy injection.

\section{Summary and outlook}

We have presented an updated analysis of the angular distribution in $\psi(3686)\to p\bar{p}$ decay, extending beyond the traditional $1+\alpha\cos^2\theta$ parameterization by explicitly incorporating the effects of transverse beam polarization and investigating potential physical sources of asymmetry. The analysis focused on two key contributions: the interference between the resonant $\psi(3686)$ amplitude and the two-photon exchange continuum, and the background from initial-state–final-state radiation (ISR–FSR) interference. A ML fit to the 1D $\cos\theta$ distribution was performed, yielding an angular parameter $\alpha = 1.00 \pm 0.03$, which is consistent with the previous BESIII measurement. The fitted contributions from the two-photon interference and the ISR–FSR background are found to be consistent with zero within uncertainties, while their inclusion is essential for a complete description of the angular distribution. We also estimate the statistical significance of $P_t$ measurement using the generated toy MC events. A significance of $5\sigma$ is achievable with about 0.1 million events when $P_t \geq 0.2$. A significant outcome of incorporating transverse polarization is the prediction of a distinct $\sin(2\phi)$ modulation in the azimuthal distribution, which clearly deviates from a flat phase-space expectation.

Our work highlights the limitations of 1D angular analyses and underscores the importance of a full 2D $(\cos\theta, \phi)$ approach to disentangle the complex dynamics at play. The observed $\sin(2\phi)$ modulation, directly linked to transverse beam polarization, serves as a new observable sensitive to the interference between the resonant and non-resonant amplitudes. Future analyses with higher statistics should implement such a 2D angular fit. This would not only allow for a more precise and simultaneous extraction of the polar asymmetry parameter $\alpha$ and the $\phi$-modulation coefficients but also enable a direct constraint on the transverse polarization degree $P_t$ from the baryon decay data itself, providing a valuable cross-check with other calibration processes like $e^+e^-\to \mu^+\mu^-$.

Looking forward, with the continued accumulation of $\psi(3686)$ data at BESIII and the potential for dedicated runs with enhanced beam polarization, future measurements will have the sensitivity to precisely determine the helicity amplitudes governing the two-photon process, thereby offering deeper insights into the interplay between QCD-driven resonance decays and QED continuum processes in the timelike region. This comprehensive angular analysis paves the way for a more complete understanding of charmonium decays into baryon-antibaryon pairs.

\section{acknowledgements}
The work is partly supported by the National Natural Science Foundation of China(NSFC) under Grants No. 12575112.

\end{document}